\documentclass[prd,superscriptaddress,amsfonts,showpacs,showkeys,twocolumn]{revtex4-1}
\usepackage[latin9]{inputenc}
\usepackage{color}
\usepackage{amsmath}
\usepackage{amssymb}
\usepackage{esint}

\usepackage{graphicx}
\usepackage{amsmath}
\usepackage{hyperref}
\usepackage{tensor}
\usepackage{color}
\usepackage{graphics}\usepackage{epsfig}\usepackage{subfigure}
\usepackage[colorinlistoftodos]{todonotes}
\usepackage[font={footnotesize,it}]{caption}
\def\be{\begin{equation}}
\def\ee{\end{equation}}
\def\bea{\begin{eqnarray}}
\def\eea{\end{eqnarray}}

\def\0{{\sst{(0)}}}
\def\1{{\sst{(1)}}}
\def\2{{\sst{(2)}}}
\def\3{{\sst{(3)}}}
\def\4{{\sst{(4)}}}
\def\5{{\sst{(5)}}}
\def\6{{\sst{(6)}}}
\def\7{{\sst{(7)}}}
\def\8{{\sst{(8)}}}
\def\sst#1{{\scriptscriptstyle #1}}

\newcommand{\nn}{\nonumber}

\begin{document}
\title{Complexity growth rates for AdS black holes with  dyonic/ nonlinear charge/ stringy hair/ topological defects }

\author{Ali \"{O}vg\"{u}n}
\email{ali.ovgun@pucv.cl}
\homepage{http://www.aovgun.com}
\affiliation{Instituto de F\'{\i}sica, Pontificia Universidad Cat\'olica de
Valpara\'{\i}so, Casilla 4950, Valpara\'{\i}so, Chile}
\affiliation{Department of Physics and Astronomy, University of Waterloo, Waterloo, Ontario, N2L 3G1, Canada }
\affiliation{Perimeter Institute for Theoretical Physics, Waterloo, Ontario, N2L 2Y5, Canada}
\affiliation{Physics Department, Arts and Sciences Faculty, Eastern Mediterranean University, Famagusta, North Cyprus via Mersin 10, Turkey}

\author{Kimet Jusufi}
\email{kimet.jusufi@unite.edu.mk}
\affiliation{Physics Department, State University of Tetovo, Ilinden Street nn, 1200,
Tetovo, Macedonia}
\affiliation{Institute of Physics, Faculty of Natural Sciences and Mathematics, Ss. Cyril and Methodius University, Arhimedova 3, 1000 Skopje, Macedonia}

\date{\today }

\begin{abstract}

In a seminal paper by Brown \textit{et al.} [Phys.\ Rev.\ Lett.\  {\bf 116}, no. 19, 191301 (2016)] a new conjecture was proposed, namely it was argued that the quantum complexity of a holographic state is equal to action of a Wheeler-DeWitt patch in the late time limit suggesting that the fastest computer in nature are the black holes. Motivated by this conjecture, in the present paper, we study the action growth rate for different types of black holes such as dyonic, nonlinear charge, stringy hair, black hole with a global monopole and a cosmic string. In general we find that action growth rates of the Wheeler-DeWitt patch is finite for these black holes at the late time approach and satisfy the Lloyd bound on the rate of quantum computation. Furthermore, in the case of a charged as well as the neutral black hole with a global monopole and a conical defect we show that the form of the Lloyd bound relation remains unaltered but the energy is modified due to the nontrivial global topology of the spacetime.
\end{abstract}

\maketitle


\section{Introduction}

Nature of humans is to always want to peel back the layers of the unknown. We look up into the sky and try to understand many unknowns from the beginning of the universe to smallest particles. Are there any limitation of our knowledge? We need more powerful helping tools to gain more information about the unexplored nature. 

Today, one of the biggest challenges that humans have is to build a quantum computer that is the topics of theoretical computer science with the helps of mathematics, namely computational complexity theory which nowadays motivates theoretical physicist \cite{qc1,qc2,qc3,qc4,b1,b2,Kempf:1994su,b3,Brown:2017jil}. Quantum complexity theory is used to shed some lights on the problems that quantum computers can solve. 

Since Maldacena enunciated the AdS/CFT correspondence which is also known as a gauge/gravity duality \cite{Maldacena:1997re}, holographic duality gains more interest \cite{Marolf:2013dba,m1,m2,m3,m4,m5,Almheiri:2012rt,Sonner:2013mba,Chernicoff:2013iga,Sinamuli:2016rms,Sinamuli:2017rhp,Sachs:2017exo,Giddings:2011ks,VanRaamsdonk:2010pw,Alishahiha:2015rta,Giddings:2012bm,Bhattacharyya:2016hbx}. Black holes are formed after the gravitational collapse in Anti-de Sitter (AdS) space-time which is dual to thermally conformal field theory. Maldecena and Susskind have made first relation between the 
Einstein-Podolsky-Rosen (EPR) correlation with the Einstein-Rosen bridge (also known as wormholes) \cite{Maldacena:2013xja}. In the other words, they have tried to connect quantum mechanics with gravity which is called ER=EPR relation. The importance of this relation is to allow the communications between Alice and Bob from opposite sides of the wormhole \cite{Susskind:2016jjb,Susskind:2017nto}.

Recently Brown et al.  have argued that the number of quantum gates, which is proportional to quantum complexity, is dual to the size of the Einstein-Rosen bridge for AdS black holes in the dual boundary CFT \cite{b1}.  
In particular this suggest that quantum complexity help us to understand better the black hole physics especially holographic duality and information paradox \cite{Polchinski:2016hrw}. On the other hand, it supports to build quantum computers. This conjecture is known as ``complexity=action'' (CA), namely the complexity of holograpic state is equal to action of a Wheeler-DeWitt patch  (on-shell action evaluated on a bulk region) in the late time limit. When finding the action growth rate, the main difficulties is to find the contribution of boundary terms \cite{b2,b3}. 

Very recently, many papers in the literature have been devoted to this conjecture to check the action growth rate and the relation between Lloyd bound \cite{lloyd} in different types of black holes whether it is valid or not \cite{Alishahiha:2015rta,Lehner:2016vdi,Swingle:2016var,Carmi:2016wjl,Carmi:2017jqz,Couch:2016exn,Chapman:2016hwi,Cai:2016xho,Roberts:2016hpo,Brown:2017jil,Ben-Ami:2016qex,Yang:2016awy,Reynolds:2016rvl,Pan:2016ecg,Barbon:2015ria,Cai:2017sjv,Chapman:2017rqy,Alishahiha:2017hwg,Kim:2017lrw,Barbon:2015soa,Huang:2016fks,Czech:2017ryf,Zangeneh:2017tub,Guo:2017rul,Abad:2017cgl,Cottrell:2017ayj,Banerjee:2017qti,Chemissany:2016qqq,Reynolds:2017lwq,Wang:2017uiw,Dvali:2016lnb,Vanchurin:2016met,Qaemmaqami:2017lzs,HosseiniMansoori:2017tsm,Ghodrati:2017roz,Nagasaki:2017kqe,Sebastiani:2017rxr,Murata:2017rbp,Susskind:2017ney,Miao:2017quj,Swingle:2017zcd,Bao:2017qmt,Fu:2016xaa,An:2018xhv}. Furthermore, Lehner et al. provide new method to find the growth rate of the action using null boundaries \cite{Lehner:2016vdi}. In this paper we shall follow the original method proposed by Brown et al. in Ref.\cite{b2}.

\begin{figure}[!ht]
\centering
\includegraphics[width=3.5in]{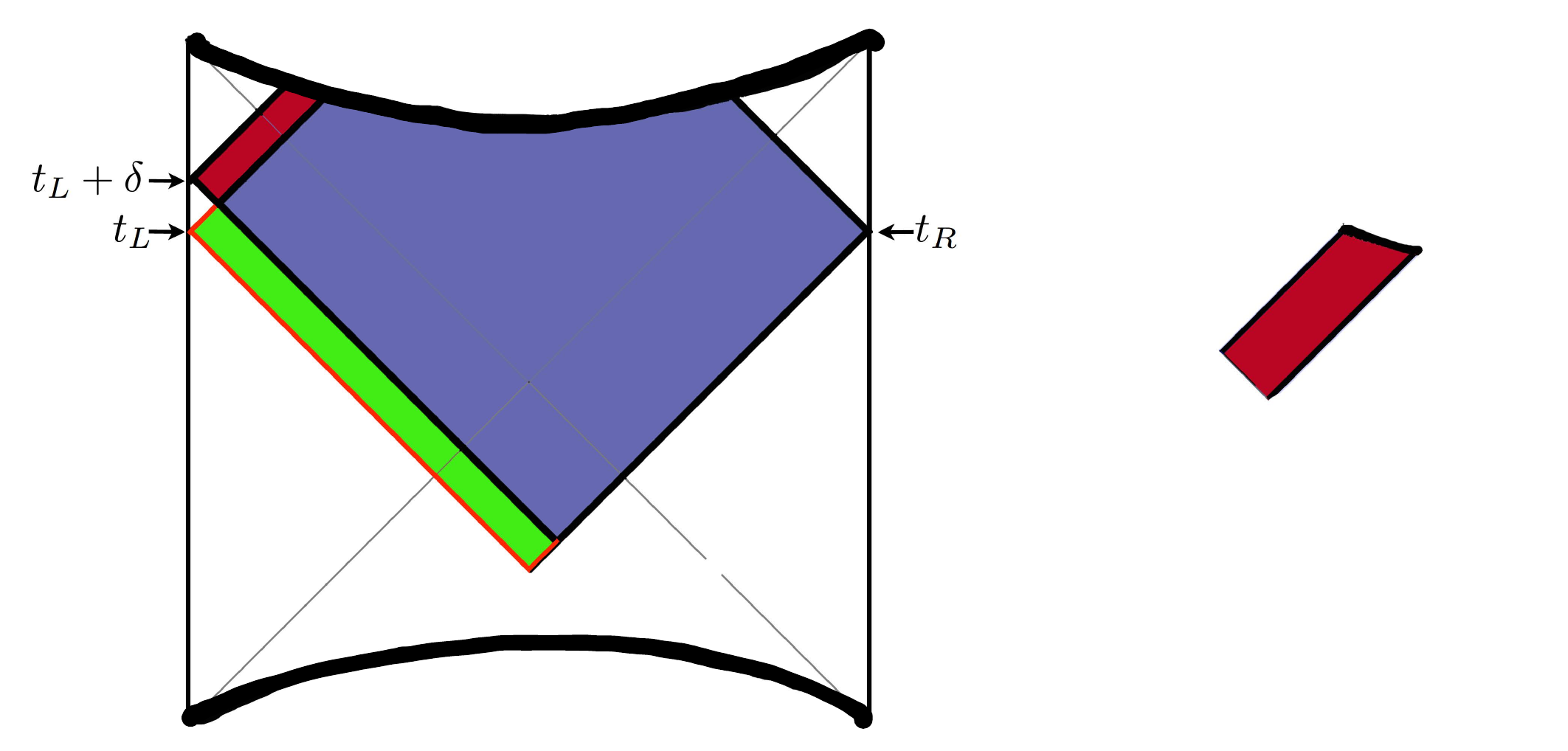}
\caption{For AdS black hole when the time increases, the WDW patch gains a  red slice and it loses a green slice for a pair of times ($t_{L}$,$t_{R}$). The figure is taken in Ref. \cite{b2}. } \label{fig1}
\end{figure}

The relation of complexity $\mathcal{C}$ with the spatial volume $V$ which is for the Einstein-Rosen bridge is given firstly by Stanford and Susskind as follows  \cite{Stanford:2014jda}:

\begin{eqnarray}
\mathcal{C}\sim\frac{V}{G l}.
\end{eqnarray}
Note that the $l$ is the anti de-Sitter (AdS) radius and the $G$ is Newton's constant.
Susskind et al. led the way through several works in which the duality of CA has been shown to be related with the action of the
black hole in the Wheeler-DeWitt patch in a following way \cite{b1,b2,b3}: 
\begin{eqnarray}
\mathcal{C}=\frac{A}{\pi\hbar}.
\end{eqnarray}

An important effect of CA duality is to show the bounded quantum complexity growth rate which is known as Lloyd bound \cite{lloyd}:

\begin{eqnarray}
\frac{d\mathcal{C}}{dt}\leq\frac{2E}{\pi\hbar},
\end{eqnarray}
with the average energy density $E$.

 Moreover, the surface of the wormhole where there is linearly growing patch provides the relation $ d\mathcal{C}/dt \approx T S $
for $t >> 1/T$, in which T is the temperature and S stands for the entropy of the black hole. This formula also gives us a hint of a growth rate of qubits so that the contact between the quantum complexity and quantum information theory can be shown \cite{hartman,Susskind:2014rva}.

There is now a growing consensus that CA duality is in the same chain of quantum complexity and action growth. In a short period of time, the seminal paper of CA duality inspired many other authors resulting with many research papers dedicated to this problematic  \cite{An:2018xhv,Swingle:2017zcd,Miao:2017quj,Sebastiani:2017rxr,Nagasaki:2017kqe,Ghodrati:2017roz,Qaemmaqami:2017lzs,Wang:2017uiw,Reynolds:2017lwq,Guo:2017rul,Alishahiha:2017hwg,Cai:2017sjv,Pan:2016ecg,Cai:2016xho}. Some exact results of the action growth rate are given as follows:
\begin{align}
\hbox{neutral BH :\quad}&\frac{\mathrm{d}\mathcal{A}}{\mathrm{d}t}=2M;\label{eq:neutralexact}\\
\hbox{rotating BH :\quad}&\frac{\mathrm{d}\mathcal{A}}{\mathrm{d}t}=\left[(M-\Omega J)_+-(M-\Omega J)_-\right];\label{eq:rotatingexact}\\
\hbox{charged BH :\quad}&\frac{\mathrm{d}\mathcal{A}}{\mathrm{d}t}=\left[(M-\mu Q)_+-(M-\mu Q)_-\right].\label{eq:chargedexact}
\end{align}
Here $\pm$  stands for the outer and inner horizons of the black hole.

In this paper, our goal is to check the validity of the complexity by calculating the action growth rate for the dyonic AdS black hole, AdS black hole with nonlinear source, AdS black hole with stringly hair, AdS black hole with global monopoles and cosmic strings. To do so, we calculate first the boundary term of the action and then the bulk action. Afterwards, we obtain the result of total action growth which is related to expected results and we carry out the main calculations of this paper to obtain the Lloyd bound. Then we compare these results with a original paper of Brown et al. \cite{b2}. It will be interesting to see the differences between the calculations of complexity of dyonic AdS black hole, AdS black hole with nonlinear source, AdS black hole with stringly hair, AdS black hole with global monopoles and cosmic strings from the original RN-AdS black hole case. The CA conjectures give us good results to understand the quantum computation. Note that in late time approximation, complexity grows linearly in time. Moreover, main contributions come from the patch from behind the horizon, known as WdW patch shown in Fig.  \ref{fig1}. It is supposed that, this linear action growth generates  the interaction between quantum states. Hence, one can consider the derivation of the time only depends on the mass/energy of the black hole/quantum states which saturates the Lloyd bound on the growth of the complexity \cite{b1,b2,b3,Reynolds:2017lwq,Reynolds:2017lwq}.

In the present article we wish to compute the action growth of the dyonic/nonlinear charge, stringly hair and global monopoles black holes in four-dimensions. Our work is organized as follows: in section II the action growth rate of the black holes with dyonic charge are discussed. In section III, we compute the complexity and action growth rate of the black holes with nonlinear charge. Then we repeat calculations for the black hole stringy hair in section IV. Then in section V, we compute the action growth rate of the charged RNAdS black hole with a global monopole. Finally in Section VI, we discuss RNAdS black hole with a conical defect. Finally, we conclude our work in Section VII.

\section{Action Growth Rate of the Dyonic Ads black holes}

Here, we consider the following action for the dyonic charged AdS black hole: \cite{Dutta:2013dca,Chaturvedi:2014vpa}

\begin{equation}
\mathcal{A}=\frac{1}{16\pi }\int d^{d}x\sqrt{-g}\left[ \mathcal{R}%
-2\Lambda -F^{\mu \nu }F_{\mu \nu }\right] ,  \label{Action}
\end{equation}%
where $\mathcal{R}$ is the Ricci scalar and $\Lambda$ stands for the
cosmological constant. It is noted that $F_{\mu \nu }$ is the electromagnetic field tensor. The electromagnetic tensor is modified to give a dyonic property, where the magnetic charge appears.  The electromagnetic 4-potential is
\begin{equation}
A=\left( -\frac{q_E}{r_-}+\frac{q_E}{r_+}\right)dt +(q_M cos\theta)d\phi
\end{equation}
in which $q_{E}$ and $q_{M}$ are respectively, electric and magnetic
charges.
The metric function yields
\begin{equation}
f(r)=1-\frac{\Lambda r^{2}}{3}-\frac{2m}{r}+\frac{q_{E}^{2}+q_{M}^{2}}{%
r^{2}},
\end{equation}

The spacetime is given by
\begin{equation}
ds^{2}=-f(r)dt^{2}+\frac{dr^{2}}{f(r)}+r^{2}d\Omega^{2},  \label{Metric2}
\end{equation}%
in which $d\Omega^{2}$ is the line element of a $(2)$-dimensional
hypersurface.  

To calculate the action growth rate \cite{b2}, we use the action of bulk and the boundary terms as follows:

\begin{eqnarray}\notag
\mathcal{A} & = & \mathcal{A}_{bk}+\mathcal{A}_{bd}\nn\label{action of LND}\\\notag
& = & \frac{1}{16\pi}\int d^{4}x\sqrt{-g}\left[ \mathcal{R}%
-2\Lambda -F^{\mu \nu }F_{\mu \nu }\right] \\
&+&\frac{1}{8\pi }\int_{\partial M}d^{3}x\sqrt{-h}K,
\end{eqnarray}
where $h$ is the induced metric for hypersurface and $K$ is the trace of the extrinsic curvature. For the dyonic AdS black hole, the action growth of the bulk is calculated as follows:

\begin{eqnarray}
\frac{\mathrm{d}\mathcal{A}_{\mathrm{bk}}}{\mathrm{d}t}=\frac{\Omega_{2}}{16\pi }\int_{r_{-}}^{r_{+}}r^{2}\left[-\frac{6}{l^2}-F^2\right]dr \nn \\ = -\frac{\Omega_{2}}{8\pi l^2}(r_+^{3}-r_-^{3})-\frac{Q^2}{2}(r_+^{-1}-r_-^{-1}),
\end{eqnarray}
where $Q^2=q_E^2 +q_M^2$.
The extrinsic curvature with the metric is 
\begin{eqnarray}
K=\frac{1}{r^2}\frac{\partial}{\partial r} \left(r^2 \sqrt{f(r)}\right)=\frac{2}{r}\sqrt{f}+\frac{f'}{2\sqrt{f}}.
\end{eqnarray}

Second we find the contribution from the YGH surface term within WDW patch at late time approximation as follows:
\begin{eqnarray} 
\frac{\mathrm{d}\mathcal{A}_{\mathrm{bd}}}{\mathrm{d}t}=\frac{1}{8\pi}\int_{r_{-}}^{r_{+}}d^{3}x\sqrt{-h}\,K.
\end{eqnarray}

Resulting with
\begin{eqnarray}
\frac{\mathrm{d}\mathcal{A}_{\mathrm{bd}}}{\mathrm{d}t}&=&\frac{\Omega_{2}}{8\pi}\left[r^{2}\sqrt{f}\left(\frac{2}{r}\sqrt{f}+\frac{f'(r)}{2\sqrt{f}}\right)\right]_{r_-}^{r_+}\\
&=&\frac{3\Omega_{2}}{8\pi l^2}(r_+^{3}-r_-^{3})+\frac{Q^2}{2}(r_+^{-1}-r_-^{-1})+\frac{2\Omega_{2}}{8\pi }(r_+ -r_-) \nn
\end{eqnarray}

Hence we obtain the total growth rate of action for dyonic AdS black hole is
\begin{align}
\frac{\mathrm{d}\mathcal{A}}{\mathrm{d}t}=\frac{2\Omega_{2}}{8\pi}\left(r_+ -r_- +\frac{r_+^{3}-r_-^{3}}{l^2}\right).
\end{align}
We rewrite above result in more compact way 
\begin{align}\label{eq:RNAdSD}
\frac{\mathrm{d}\mathcal{A}}{\mathrm{d}t}=Q^2\left(\frac{1}{r_-}-\frac{1}{r_+}\right).
\end{align}

by using the mass $M$:
\begin{align}\label{eq:M}
M=\frac{2\Omega_{2}}{16\pi}\left(r_+ +r_-+\frac{1}{l^2}\frac{r_+^{4}-r_-^{4}}{r_+ -r_-}\right),
\end{align}
and charge $Q$:
\begin{align}\label{eq:Q2}
Q^2=\frac{2\Omega_{2}}{8\pi}r_+ r_- \left(1+\frac{1}{l^2}\frac{r_+^{3}-r_-^{3}}{r_+ -r_-}\right).
\end{align}
The growth rate of action within WDW patch for dyonic-AdS black hole at late time approximation is obtained similarly the seminal paper of Brown  et al. \cite{b1,b2}
\begin{align}\label{eq:RNAdSDpm}
\frac{\mathrm{d}\mathcal{A}}{\mathrm{d}t}=(M-\mu_+Q)-(M-\mu_-Q),
\end{align}
in which, $\mu_-=Q/r_-$ and $\mu_+=Q/r_+$ stand for the chemical potentials at inner/ outer horizons. Moreover it satisfies the Lloyd bound \cite{lloyd}.

 In particular the quantum complexity growth rate is bounded by 
\begin{eqnarray}
\frac{\mathrm{d}\mathcal{C}}{\mathrm{d}t}\leq\frac{2E}{\pi\hbar},
\end{eqnarray}
where $E$ is the average energy of the quantum state relating to
the ground state. Hence, the dyonic charge has regular impact on the complexity similarly charged AdS black hole.

\bigskip
\section{Action Growth Rate of the black holes with a nonlinear source}

 In this section, we use the black hole with a nonlinear source to calculate the complexity. The total action with bulk term using the  Einstein-power Maxwell invariant (PMI) gravity and boundary term is given by \cite{PMIpapers2}
\begin{eqnarray}\notag
\mathcal{A} & = & \mathcal{A}_{bk}+\mathcal{A}_{bd}\nn\label{action of LND1}\\\notag
& = & \frac{1}{16\pi }\int_{M}d^{4}x\sqrt{-g}\left( R+\frac{6}{l^{2}}%
+\mathcal{L}_{PMI}\right)\\
&+&\frac{1}{8\pi G}\int_{\partial M}d^{3}x\sqrt{-h}K,
\end{eqnarray}
where $\mathcal{L}_{PMI}=(-\mathcal{F})^{s}$ and $\mathcal{F}=F_{\mu \nu
}F^{\mu \nu }$. It is noted that, we use the special case of $s=3/2$.

Using this PMI gravity, the following spherically symmetric spacetime is obtained as follows \cite{PMIpapers2,PMIpapers3,PMIpapers4,PMIpapers6,PMIpapers7}:
\begin{equation}
ds^{2}=-f(r)dt^{2}+\frac{dr^{2}}{f(r)}+r^{2}d\Omega^{2},
\label{Metric3}
\end{equation}
where $d\Omega^{2}$ is for the standard element on $S^{2}$, and the metric function is
\begin{eqnarray}
f(r) &=&1+\frac{r^{2}}{l^{2}}-\frac{2m}{r}+\frac{2^{3/2}q^3 \ln(r)}{r}, \label{metfunction}
\end{eqnarray}
Now we calculate straightforwardly the action growth of the bulk:
\begin{eqnarray}
\frac{\mathrm{d}\mathcal{A}_{\mathrm{bk}}}{\mathrm{d}t}=\frac{\Omega_{2}}{16\pi }\int_{r_{-}}^{r_{+}}r^{2}\left[\,{\frac {-24\,{r}^{6}+{q}^{3}{l}^{2}}{4{l}^{2}{r}^{6}}}\right]dr \nn \\ =\,{\frac {\Omega_{2}\,{{\it r_{-}}}^{3}}{8\pi\,{l}^{2}}}-\,{\frac {
\Omega_{2}\,{{\it r_{+}}}^{3}}{8\pi\,{l}^{2}}}-{\frac {\,{q}^{3}}{6\,{
{\it r_{+}}}^{3}\pi}}+{\frac {\,{q}^{3}}{6\,\pi\,{{\it r_{-}}}^{3}}}.
\end{eqnarray}

Then we obtain the extrinsic curvature:
\begin{eqnarray}
K=\frac{1}{r^2}\frac{\partial}{\partial r} \left(r^2 \sqrt{f(r)}\right)=\frac{2}{r}\sqrt{f}+\frac{f'}{2\sqrt{f}},
\end{eqnarray}
to calculate the YGH surface term within WDW patch at late time approximation as follows:
\begin{widetext}
\begin{align}
\frac{\mathrm{d}\mathcal{A}_{\mathrm{bd}}}{\mathrm{d}t}&=\frac{\Omega_{2}}{8\pi}\left[r^{2}\sqrt{f}\left(\frac{2}{r}\sqrt{f}+\frac{f'(r)}{2\sqrt{f}}\right)\right]_{r_-}^{r_+}\nonumber\\
&=\,{\frac {\Omega_{2}\, \left(  \left( {\it r_{-}}\,{l}^{2}+3\,{{\it r_{-}}}^
{3} \right) \ln  \left( {\it r_{-}} \right) + \left( -{\it r_{-}}\,{l}^{2}-3
\,{{\it r_{-}}}^{3} \right) \ln  \left( {\it r_{+}} \right) -{\it r_{-}}\,{l}^{
2}+{\it r_{+}}\,{l}^{2}-{{\it r_{-}}}^{3}+{{\it r_{+}}}^{3} \right) }{16\pi\,
 \left( \ln  \left( {\it r_{-}} \right) -\ln  \left( {\it r_{+}} \right) 
 \right) {l}^{2}}}
\end{align}
\end{widetext}

Note that we use $f(r_{\pm})=0$.
Using the late time approximation, it can be read off the total growth rate of action for dyonic AdS black hole within WDW patch as follows:

\begin{widetext}
\begin{align}
\frac{\mathrm{d}\mathcal{A}}{\mathrm{d}t}=\,{\frac { \left( -2\,\sqrt {2} \left( {\it r_{-}}\,{l}^{2}-{\it r_{+}}\,
{l}^{2}+{{\it r_{-}}}^{3}-{{\it r_{+}}}^{3} \right) l \left( \ln  \left( {
\it r_{-}} \right) -\ln  \left( {\it r_{+}} \right)  \right) ^{2} \right) ^{
2/3} \left( -{\it r_{+}}+{\it r_{-}} \right) }{2{l}^{2} \left( -\ln  \left( {
\it r_{-}} \right) +\ln  \left( {\it r_{+}} \right)  \right) ^{2}{\it r_{-}}\,{
\it r_{+}}}}
.
\end{align}
\end{widetext}
To write it in more compact way, we obtain the mass $m$
\begin{align}\label{eq:M2}
m=\,{\frac { \left( {\it r_{+}}\,{l}^{2}+{{\it r_{+}}}^{3} \right) \ln 
 \left( {\it r_{-}} \right) -{\it r_{-}}\,\ln  \left( {\it r_{+}} \right) 
 \left( {l}^{2}+{{\it r_{-}}}^{2} \right) }{ 2\left( \ln  \left( {\it r_{-}}
 \right) -\ln  \left( {\it r_{+}} \right)  \right) {l}^{2}}},
\end{align}
and total charge $q$
\begin{equation}\label{eq:Q}
q={\frac {\sqrt [3]{-2\sqrt {2} \left( {\it r_{-}}{l}^{2}-{\it r_{+}}
{l}^{2}+{{\it r_{-}}}^{3}-{{\it r_{+}}}^{3} \right) l \left( \ln  \left( {
\it r_{-}} \right) -\ln  \left( {\it r_{+}} \right)  \right) ^{2}}}{2l
 \left( \ln  \left( {\it r_{-}} \right) -\ln  \left( {\it r_{+}} \right) 
 \right) }}
\end{equation}

Hence, the total action growth rate for dyonic AdS black hole becomes 
\begin{align}\label{eq:RNAdSDpm2}
\frac{\mathrm{d}\mathcal{A}}{\mathrm{d}t}=(m-\mu_+q)-(m-\mu_-q).
\end{align}
Here also it reduces to original charged AdS black hole case and satisfy the Lloyd bound.
 In particular the quantum complexity growth rate is bounded by 
\begin{eqnarray}
\frac{\mathrm{d}\mathcal{C}}{\mathrm{d}t}\leq\frac{2E}{\pi\hbar},
\end{eqnarray}
where $E$ is the average energy of the quantum state relating to
the ground state. Therefore, the black hole with a nonlinear source reduces to normal charged AdS black hole, when nonlinear term is gone.

\section{Action growth rate of the AdS black holes with stringly hair}

We consider an action in which gravity is coupled to electrodynamic
field as \cite{str,str2}
\begin{equation}
\mathcal{A}=\frac{1}{16\pi}\int{d^{4}x\sqrt{-g}\left[\mathcal{R}-2\Lambda+L(F)+||J||\right]},\label{Act}
\end{equation}
where the string field is $J=H(1-\Omega_{2})$ with $H=-e^{2 \sigma} \Delta \sigma$, Ricci scalar curvature is $\mathcal{R}$ and $\Lambda$
stands for the cosmological constant. $L(F)$ is the Lagrangian of linear
electrodynamics field given by $L(F)=-\frac{1}{4}F^{2}$ where $F^{2}=F_{\mu\nu}F^{\mu\nu}$,
with $F_{\mu\nu}=\partial_{\mu}A_{\nu}-\partial_{\nu}A_{\mu}$ is
the electromagnetic field tensor. $F=\frac{Q}{r^{2}}$.

The spacetime of the black hole with stringy hair (BHSH) is recently
found by Boos and Frolov \cite{Boos:2017pyd}:
\begin{align}
\mathrm{d}s^{2} & =-f(r)\mathrm{d}t^{2}+f^{-1}(r)\mathrm{d}r^{2}+\mathrm{d}\omega_{0}^{2},\label{metric1}\\
 & d\omega_{0}^{2}=r^{2}e^{2\sigma}\left(d\theta^{2}+\sin^{2}\theta\,\mathrm{d}\varphi^{2}\right),
\end{align}
where $\sigma$ is constant which depends on $\theta$ and $\phi$, but in this paper we chose it as a $\sigma_0$.
The interesting feature of the spacetime is that the metric is warped and distorted with $d\omega_0^{2}$. The radius of the black hole is located at $f(r_+)= 0$, where the metric function is 
\begin{equation}
f(r)=1-\frac{2M}{r}+\frac{Q^{2}}{r^{2}}-r^2 \Lambda.
\end{equation}
It is noted that $M$ is a mass of the black hole and the charge of the black hole is $Q$. 
To study the complexity on the black hole with stringly hair, we write the total action with bulk term and boundary term as follows:

\begin{eqnarray}\notag
\mathcal{A} & = & \mathcal{A}_{bk}+\mathcal{A}_{bd}\nn\label{action of LND3}\\\notag
& = & \frac{1}{16\pi}\int d^{4}x\sqrt{-g}\left[ \mathcal{R}%
-2\Lambda -F^{\mu \nu }F_{\mu \nu }\right] \\
&+&\frac{1}{8\pi}\int_{\partial M}d^{3}x\sqrt{-h}K,
\end{eqnarray}

Then we find the action growth of the bulk

\begin{eqnarray}
\frac{\mathrm{d}\mathcal{A}_{\mathrm{bk}}}{\mathrm{d}t}=\frac{\Omega_{2}}{16\pi }\int_{r_{-}}^{r_{+}} e^{2\sigma_0}r^{2}\left[-\frac{6}{l^2}-F^2\right]dr \nn \\ = e^{2\sigma_0}\left[ -{\frac {Q^{2}}{2\,{\it r_{+}}}}+{\frac {{Q}^{2}}{2\,{\it r_{-}}}}+\,
{\frac {{{\it \Omega_{2} r_{-}}}^{3}}{8\pi{l}^{2}}}-\,{\frac {{{\it \Omega_{2} r_{+}}}^{3}}{8\pi{l}^{2}
}}\right]
\end{eqnarray}
and we calculate the YGH surface term within WDW patch at late time approximation resulting with
\begin{align}
\frac{\mathrm{d}\mathcal{A}_{\mathrm{bd}}}{\mathrm{d}t}&=\frac{\Omega_{2}}{8\pi}\left[r^{2}e^{2\sigma_0}\sqrt{f}\left(\frac{2}{r}\sqrt{f}+\frac{f'(r)}{2\sqrt{f}}\right)\right]_{r_-}^{r_+}\nonumber\\
&=\frac{2\Omega_{2} e^{2\sigma_0}}{8\pi l^2}(r_+^{3}-r_-^{3})+\frac{Q^2 e^{2\sigma_0}}{2}(r_+^{-1}-r_-^{-1}) \nn
\\ &+\frac{2\Omega_{2} e^{2\sigma_0}}{8\pi }(r_+ -r_-).
\end{align} 

Finally we can write the total action growth rate as:
\begin{align}
\frac{\mathrm{d}\mathcal{A}}{\mathrm{d}t}=\frac{2\Omega_{2}e^{2\sigma_0}}{8\pi}\left(r_+ -r_- +\frac{r_+^{3}-r_-^{3}}{l^2}\right).
\end{align}
and after considering  $\sigma_0=0$, the growth rate of action of the black hole with stringly hair  reduces to
\begin{align}
\frac{\mathrm{d}\mathcal{A}}{\mathrm{d}t}=(M-\mu_+Q)-(M-\mu_-Q).
\end{align}
 In particular the quantum complexity growth rate is bounded by 
\begin{eqnarray}
\frac{\mathrm{d}\mathcal{C}}{\mathrm{d}t}\leq\frac{2E}{\pi\hbar},
\end{eqnarray}
where $E$ is the average energy of the quantum state relating to
the ground state. It is noted that the effect of the stringly hair can be 

\section{Action growth rate of charged AdS black hole with a global monopole
}

A global monopole is an interesting object with a wide range of physical implications in the context of gravity theory as well as quantum theory. It is speculated that these objects can arise during the phase transition of a system composed by a self-coupling scalar triplet $\phi^{a}$ in the early universe. In particular, the simplest model of such scenario can be studied by the following Lagrangian density \cite{gb1,gb2}
\begin{equation}\label{gbmetric}
\mathcal{L}_{GB}=-\frac{1}{2}\sum_a g^{\mu\nu}\partial_{\mu}\phi^{a} \partial_{\nu}\phi^{a}-\frac{\lambda}{4}\left(\phi^{2}-\eta^{2}\right)^{2},
\end{equation}
with $a=1, 2, 3$, with $\lambda$ being the self-interaction term, $\eta$ is known as the scale of a gauge-symmetry breaking. The field of such a system is given by 
\begin{equation}
\phi^{a}=\frac{\eta h(r) x^{a}}{r},
\end{equation}
where
\begin{equation}
x^{a}=\left\lbrace r \sin\theta \, \cos\varphi, r \sin\theta \,\sin\varphi,r \cos\theta \,\right\rbrace,
\end{equation}
such that $\sum_a x^{a}x^{a}=r^{2}$. Using the field equations and the relation for $\phi^a$ one can show that the problem reduces to a single equation for $h(r)$ given as \cite{gb1}
\begin{equation}
f h''+\left[\frac{2f}{r}+\frac{1}{2f} (f^2)' \right]h'-\frac{2h}{r^2}-\lambda \eta^2 h \left(h^2-1\right)=0.
\end{equation}

Interestingly, outside the core in the large limit approximation one can take  $h(r)\to 1$, with the energy-momentum tensor given by the following relations $T^{t}_{t}=T^{r}_{r}\simeq \eta^{2}/r^2$ and $T^{\theta}_{\theta}=T^{\varphi}_{\varphi}=0$. The global monopole metric (also known as Barriola-Vilenkin metric) with cosmological constant is given as follows \cite{gb1,gb2}
\begin{equation}
ds^2=-f(r)dt^2+\frac{dr^2}{f(r)}+r^2\left(d\theta^2+\sin^2\theta d\varphi^2 \right),
\end{equation}
where 
\begin{equation}
f(r)=1-8 \pi \eta^2 -\frac{2M}{r}+\frac{Q^2}{r^2}+\frac{ r^2}{l^2}.
\end{equation}

In the last expression $M\approx M_{core}$ denotes the global monopole core mass, with $M_{core}\approx \lambda^{-1/2} \eta$, note that for a typical grand unification scale $\eta=10^{16}$ GeV. The total action if our system reads 
\begin{eqnarray}\notag
\mathcal{A} &=& \frac{1}{16\pi}\int{d^{4}x\sqrt{-g}\left(\mathcal{R}-2\Lambda-F_{\mu \nu}F^{\mu \nu}\right)}\\
&+&\int d^4x \sqrt{-g} \,\mathcal{L}_{GB}+\frac{1}{8\pi}\int_{\partial M}d^{3}x\sqrt{-h}K.
\end{eqnarray}

Introducing the following coordinate transformation into the metric (47) given as \cite{gb3}
\begin{eqnarray}\notag
t & \to & (1-8 \pi \eta^2)^{-1/2}t,\\\notag
r & \to & (1-8 \pi \eta^2)^{1/2}r, \\\notag
M & \to & (1-8 \pi \eta^2)^{-3/2}M, \\\notag
Q & \to &  (1-8 \pi \eta^2)^{-1}Q,\notag
\end{eqnarray}
we find the following result
\begin{equation}
ds^2=-f(r)dt^2+\frac{dr^2}{f(r)}+(1-8 \pi \eta^2) r^2\left(d\theta^2+\sin^2\theta d\varphi^2 \right)
\end{equation}
in which
\begin{equation}
f(r)=1-\frac{2M}{r}+\frac{Q^2}{r^2}+\frac{ r^2}{l^2}.
\end{equation}

We shall calculate now the contribution from the bulk action as
\begin{eqnarray}\notag
\frac{\mathrm{d}\mathcal{A}_{bk}}{\mathrm{d}t}&=&\int  \mathrm{d}^{4}x{\sqrt{-g} \left[\frac{1}{16\pi} \left(\mathcal{R}\text{ }-2\Lambda -F_{\mu \nu}F^{\mu \nu}\right) +\mathcal{L}_{GB}\right]}\\
 &=& \frac{\Omega_2(1-8 \pi \eta^2)}{16 \pi }\int_{r_{-}}^{r_{+}} r^2 \Xi(r,l,Q,\eta)\, \mathrm{d}r,
\end{eqnarray}
with
\begin{equation}
\Xi= -{\frac {16\,\pi\,{\eta}^{2}{l}^{2}+96\,\pi\,{\eta}^{2}{r}^{2}-12\,{r}
^{2}}{ \left( 8\,\pi\,{\eta}^{2}-1 \right) {r}^{2}{l}^{2}}}+\frac{6}{l^2}+\frac{2 Q^2}{r^4}   
\end{equation}
with the Ricci scalar given by
\begin{equation}
\mathcal{R}=-{\frac {16\,\pi\,{\eta}^{2}{l}^{2}+96\,\pi\,{\eta}^{2}{r}^{2}-12\,{r}
^{2}}{ \left( 8\,\pi\,{\eta}^{2}-1 \right) {r}^{2}{l}^{2}}}.
\end{equation}

Hence the action growth rate gives
\begin{eqnarray}\notag
\frac{\mathrm{d}\mathcal{A}_{bk}}{\mathrm{d}t}&=& Q^2(1-8 \pi \eta^2) \left( \frac{1}{2r_-}-\frac{1}{2r_+}\right)-\frac{r_{+}^3-r_{-}^3}{2l^2}\\
&+&\frac{4 \pi \eta^2 (r_{+}-r_{-})(l^2+r_+^2+r_+ r_{-}+r_{-}^2)}{l^2}
\end{eqnarray}

On the other hand, first we find that the extrinsic curvature remains unchanged due to the presence of a global monopole, namely we find
\begin{eqnarray}
K=\frac{2 \sqrt{f(r)}}{r}+\frac{f'(r)}{2 \sqrt{f(r)}}.
\end{eqnarray}

Thus, the contribution from the YGH surface term yields
\begin{eqnarray}\notag
\frac{\mathrm{d}\mathcal{A}_{bd}}{\mathrm{d}t}&=&\frac{(1-8 \pi \eta^2)\Omega_2}{8 \pi}\left[r^2\sqrt{f(r)} K  \right]_{r_{-}}^{r_{+}}\\\notag
&=&\frac{3(1-8 \pi \eta^2)}{2l^2}(r_+^3-r_-^3)+(1-8 \pi \eta^2)(r_+-r_{-})\\
&+&(1-8 \pi \eta^2)Q^2\left(\frac{1}{2r_+}-\frac{1}{2r_{-}}\right).
\end{eqnarray}

For the total action growth rate we find, 
\begin{eqnarray}\notag
\frac{d\mathcal{A}}{dt} & = & \frac{d\mathcal{A}_{bk}}{dt}+\frac{d\mathcal{A}_{bd}}{dt}\nn\\\notag
 & = & (1-8 \pi \eta^2)(r_+-r_{-})+\frac{[3(1-8 \pi \eta^2)-1](r_{+}^3-r_{-}^3)}{2l^2}\\
 &+& \frac{4 \pi \eta^2 (r_{+}-r_{-})(l^2+r_+^2+r_+ r_{-}+r_{-}^2)}{l^2}.
\end{eqnarray}

From $f(r_{+})=0$, we find $M$ given by
\begin{eqnarray}
M=\frac{r_+^4+r_+^2 l^2+Q^2l^2}{2 r_+ l^2}.
\end{eqnarray}

If we use this equation, from $f(r_{-})=0$ on the other hand we find $Q^2$, as follows
\begin{eqnarray}
Q^2=\frac{r_+ r_{-}l^2+r_+ r_{-}^3+r_{+}^2 r_{-}^2+r_{+}^3 r_{-}}{l^2}.
\end{eqnarray}

Finally, putting all these results together we obtain
\begin{equation}
\frac{\mathrm{d}\mathcal{A}}{\mathrm{d}t} =   4 \pi \eta^2(r_+-r_-)+(1-8 \pi \eta^2)\left(\frac{Q^2}{r_-} -\frac{Q^2}{r_+} \right),
\end{equation}
or 
\begin{eqnarray}\notag
\frac{\mathrm{d}\mathcal{A}}{\mathrm{d}t} & = &  \left[M+4 \pi \eta^2 r_+-(1-8 \pi \eta^2) \mu_+ Q \right]\\
&-& \left[M+4 \pi \eta^2 r_{-}-(1-8 \pi \eta^2) \mu_- Q \right].
\end{eqnarray}

It is worth noting that in the last equation the  chemical potentials on the horizons are given by $\mu_{\pm}=Q/r_{\pm}$. Furthermore if we define the ADM charge which corresponds to the charge measured at infinity, given by 
\begin{equation}
\mathcal{Q}=\frac{1}{4 \pi}\int F^{\mu \nu} \mathrm{d}^2\Sigma_{\mu \nu}=(1-8 \pi \eta^2) Q, 
\end{equation}
the above simplifies to 
\begin{equation}
\frac{d\mathcal{A}}{dt} =   \left[M+4 \pi \eta^2 r_{+}-\mu_+ \mathcal{Q}  \right]- \left[M+4 \pi \eta^2 r_{-}-\mu_{-} \mathcal{Q} \right].
\end{equation}

In the limit of Schwarzschild-AdS black hole one has $r_- \to 0$, $r_+ \to 2M$, consequently $\mu_+ Q \to 0$, and $\mu_{-}Q \to 2M$. It follows 
\begin{equation}
\frac{\mathrm{d}\mathcal{A}}{\mathrm{d}t}  \to  2M \left(1-4 \pi \eta^2 \right).
\end{equation}  

Thus, we have shown that the presence of global monopole modifies the growth rate action for the neutral as well as the charged black hole. In particular the quantum complexity growth rate is bounded by 
\begin{eqnarray}
\frac{\mathrm{d}\mathcal{C}}{\mathrm{d}t}\leq\frac{2E}{\pi\hbar}\left(1-4 \pi \eta^2 \right).
\end{eqnarray}

The last equation shows that the average energy of the quantum state relating to
the ground state $E$, is modified in the presence of a global monopole with the Lloyd bound given relation
\begin{equation}
\frac{\mathrm{d}\mathcal{C}}{\mathrm{d}t}\leq\frac{2\mathcal{E}}{\pi\hbar}.
\end{equation}  
where we have introduced the modified average energy of the quantum state given by $\mathcal{E}=(1-4 \pi \eta^2 ) E$. 

\section{Action growth rate of RNAdS black hole with a cosmic string}

The spacetime metric of a RNAdS spacetime with a cosmic string is given as \cite{cd}
\begin{equation}
ds^2=-f(r)dt^2+\frac{dr^2}{f(r)}+r^2\left[d\theta^2+(1-4  \mu)^2\sin^2\theta d\varphi^2 \right],
\end{equation}
where 
\begin{equation}
f(r)=1 -\frac{2M}{r}+\frac{Q^2}{r^2}+\frac{ r^2}{l^2}.
\end{equation}

The total action of the system  is given by
\begin{eqnarray}\notag
\mathcal{A}&=&\frac{1}{16\pi}\int{d^{4}x\sqrt{-g}\left(\mathcal{R}-2\Lambda-F_{\mu \nu}F^{\mu \nu}\right)}+\mathcal{A}_{string}\\
&+&\frac{1}{8\pi}\int_{\partial M}d^{3}y\sqrt{-h}K.
\end{eqnarray}

In which the action associated to a cosmic string without internal structure aligned in the $z$-axes can be given as follows 
\begin{equation}
\mathcal{A}_{string}=-\frac{1}{2} \int d^2\zeta \,\sqrt{-\gamma}\,\, {T^{\mu}}_{\mu} 
\end{equation}
where ${T^{\mu}}_{\mu}=2 \mu \delta(x) \delta(y)$, in which  $\mu$ is the tension of the cosmic string. Furthermore, $\zeta^a$ are coordinates on the string world-sheet. First we calculate the contribution from the bulk action which gives
\begin{equation}
\frac{\mathrm{d}\mathcal{A}_{bk}}{\mathrm{d}t}= \frac{\Omega_2(1-4 \mu)}{16 \pi }\int_{r_{-}}^{r_{+}} r^2 \left(-\frac{12}{l^2}+\frac{6}{l^2} + \frac{2Q^2}{r^4}\right) \mathrm{d}r,
\end{equation}
where the Ricci scalar is found to be
\begin{equation}
\mathcal{R}=-\frac{12}{l^2}.
\end{equation}

Note that there is zero contribution from the cosmic string action. Hence the action growth rate gives
\begin{equation}
\frac{\mathrm{d}\mathcal{A}_{bk}}{\mathrm{d}t}= Q^2\left(1-4 \mu\right)\left(\frac{1}{2 r_{-}}-\frac{1}{2 r_{+}}  \right)-\left(r_{+}^3-r_{-}^3  \right) \frac{(1-4\mu)}{2 l^2}
\end{equation}

And the contribution from the YGH surface term is 
\begin{eqnarray}\notag
\frac{\mathrm{d}\mathcal{A}_{bd}}{\mathrm{d}t}&=&\frac{(1-4 \mu)\Omega_2}{8 \pi}\left[r^2\sqrt{f(r)} K  \right]_{r_{-}}^{r_{+}}\\\notag
&=&\frac{3(1-4 \mu)}{2l^2}(r_+^3-r_-^3)+(1-4 \mu)(r_+-r_{-})\\
&+&(1-4 \mu )Q^2\left(\frac{1}{2r_+}-\frac{1}{2r_{-}}\right).
\end{eqnarray}

Then we can get the total action growth rate, 
\begin{eqnarray}
\frac{d\mathcal{A}}{dt} & = & \frac{d\mathcal{A}_{bk}}{dt}+\frac{d\mathcal{A}_{bd}}{dt}\nn\\
 & = & (1-4 \mu)(r_+-r_{-})+\frac{(1-4 \mu)(r_{+}^3-r_{-}^3)}{l^2}
\end{eqnarray}

It follows that 
\begin{eqnarray}
\frac{d\mathcal{A}}{dt} & = & (1-4\mu) \left( \frac{Q^2}{r_{-}}-\frac{Q^2}{r_{+}}\right),
\end{eqnarray}
in other words 
\begin{equation}
\frac{d\mathcal{A}}{dt}  = \left[M-(1-4\mu)\mu_{+}Q \right]-\left[M-(1-4\mu)\mu_{-}Q \right].
\end{equation}

One can introduce the ADM charge in the spacetime background of a cosmic string given by $\mathcal{Q}=(1-4\mu)Q$, in that case the above equation simplifies to 
\begin{equation}
\frac{d\mathcal{A}}{dt}  = \left[M-\mu_{+}\mathcal{Q} \right]-\left[M-\mu_{-}\mathcal{Q} \right].
\end{equation}

In the limit of Schwarzschild-AdS black hole we have $\mu_+ Q \to 0$, and $\mu_{-}Q \to 2M$. It follows 
\begin{eqnarray}
\frac{\mathrm{d}\mathcal{A}}{\mathrm{d}t} & \to & 2M \left(1-4 \mu \right).
\end{eqnarray}  

Thus, we have shown that the presence of a conical defect modifies the growth rate action for the neutral black hole as well as the charged black hole. In particular, the quantum complexity growth rate is bounded by 
\begin{eqnarray}
\frac{\mathrm{d}\mathcal{C}}{\mathrm{d}t}\leq\frac{2E}{\pi\hbar}\left(1-4 \mu \right),
\end{eqnarray}
  
When a topological defect is introduced the global spacetime topology becomes nontrivial, hence it is convenient to introduce the ADM mass in the presence of cosmic string which gives $\mathcal{M}=(1-4\mu) M$, yielding
\begin{equation}
\frac{\mathrm{d}\mathcal{A}}{\mathrm{d}t}  \to  2 \mathcal{M}.
\end{equation}

In that case, the Lloyd bound can be written as
\begin{equation}
\frac{\mathrm{d}\mathcal{C}}{\mathrm{d}t}\leq\frac{2\mathcal{E}}{\pi\hbar}.
\end{equation}  
where the average energy of the quantum state in presence of conical defects is modified as $\mathcal{E}=(1-4\mu) E$. 

\bigskip
\section{Conclusion}

In this paper,  using the ``complexity$=$action" (CA) conjecture the complexity growth rate is studied in the AdS black holes with dyonic/ nonlinear charge/ stringy hair/ topological defects.

We have investigate the boundary term of the action and the bulk action to calculate the action growth rate. Afterwards, we obtain the result of the total action growth which is related to results of the seminal paper of the Brown et al. \cite{b2,b2}  and we carry out the main calculations of this paper to obtain the Lloyd bound. Further we have explored the differences between the calculations of complexity of dyonic AdS black hole, AdS black hole with nonlinear source, AdS black hole with stringly hair, AdS black hole with global monopoles and cosmic strings from the original RN-AdS black hole case. 

For the black hole with dyonic charge, we found that the action growth rate of the black hole depend on the total charge where the dyonic charge is emerged.

On the other hand, for the black hole with nonlinear source as well as the black hole with stringly hair we find that the action growth rate reduces to the familiar charged black hole case reported in the literature. In other words, the Lloyd bound is fulfilled in all three cases.

On the other hand, in the case of RNAdS black hole with a global monopole we find that the quantum complexity growth rate is bounded by 
\begin{eqnarray}\notag
\frac{\mathrm{d}\mathcal{C}}{\mathrm{d}t}\leq\frac{2E}{\pi\hbar}\left(1-4 \pi \eta^2 \right).
\end{eqnarray}

Thus, due to the presence of a global monopole the Lloyd bound is slightly modified.  Lastly, we have used the black hole with a conical defects (cosmic string) which give us a fruitful result 
\begin{eqnarray}\notag
\frac{\mathrm{d}\mathcal{C}}{\mathrm{d}t}\leq\frac{2E}{\pi\hbar}\left(1-4 \mu \right).
\end{eqnarray}

Hence, the form of Lloyd bound  relation remain unaltered but the energy changes. This modification of the energy, however, is to be expected due to the nontrivial global topology of the spacetime when topological defects are introduced. For this reason, we have used the ADM charge, as well as the ADM mass in the total action growth rate. Another interesting way to find similar result for the complexity growth rate of the WDW patch at late time point is using the approach proposed by Lehner et al.'s method \cite{Lehner:2016vdi}.

More importantly, the CA conjectures provide interesting results to shed light on the quantum computation. Furthermore complexity grows linearly in time if one propose the late time approximation where the main contributions come from the WdW patch. Hence the linear rates of the action growth support the link between quantum states and it saturates the Lloyd bound on the growth of the complexity. 

This is another important evidence for the idea that black holes are the fastest computers and scramblers in nature. It would also be very interesting to investigate the complexity growth rate, which is link between space-time geometry and quantum entanglements \cite{Susskind:2014moa}, in different gravity theories and different geometries to understand deeply it's nature. We will leave it to our future projects.

\acknowledgments
This work was supported by the Chilean FONDECYT Grant No. 3170035 (A\"{O}). A\"{O} is grateful to the Waterloo University, Department of Physics and Astronomy, and also Perimeter Institute for Theoretical Physics for hosting him as a research visitor where part of this work was done.

\end{document}